\documentclass[12pt,a4,a4wide]{article}
\usepackage{amsmath}
\usepackage{amssymb}
\usepackage{epsfig}
\usepackage{mcite}

\def\be{\begin{equation}}
\def\ee{\end{equation}}

\newcommand{\ba}{\begin{eqnarray}}  
\newcommand{\bad}{\begin{array}{ccc}}
\newcommand{\bea}{\begin{equation} \begin{array}{c}}
\newcommand{\eea}{ \end{array} \end{equation}}
\newcommand{\ea}{\end{eqnarray}}

\newcommand{\Ord}{\ensuremath{{\cal O}}}
\newcommand{\mev}{\ensuremath{\mathrm{MeV}}}

\newcommand{\gev}{\ensuremath{\mathrm{GeV}}}

\newcommand{\eL}{\ensuremath{{\bf e}}_{\mathrm L}}

\newcommand{\eN}{\ensuremath{{\bf e}}_{\mathrm N}}
\newcommand{\eT}{\ensuremath{{\bf e}}_{\mathrm T}}
\newcommand{\ei}{\ensuremath{{\bf e}}_{\mathrm i}}
\newcommand{\eibar}{\ensuremath{{\bf \bar e}}_{\mathrm i}}
\newcommand{\pq}{\ensuremath{{\bf p}}_{q}}
\newcommand{\pl}{\ensuremath{{\bf p}}_{l^-}}
\newcommand{\vecn}{\ensuremath{{\bf n}}}
\newcommand{\PL}{\ensuremath{{P}_{\rm L}}}
\newcommand{\PN}{\ensuremath{{P}_{\rm N}}}
\newcommand{\PT}{\ensuremath{{P}_{\rm T}}}
\newcommand{\PI}{\ensuremath{{P}_{\rm i}}}
\newcommand{\ACP}{\ensuremath{{A}_{\rm CP}}}
\newcommand{\ACPL}{\ensuremath{\delta{A}_{\rm CP}^{\rm L}}}
\newcommand{\ACPN}{\ensuremath{\delta{A}_{\rm CP}^{\rm N}}}
\newcommand{\ACPT}{\ensuremath{\delta{A}_{\rm CP}^{\rm T}}}
\newcommand{\lu}{\ensuremath{{\lambda}_{u}}}

\newcommand{\shat}{\ensuremath{{\hat s}}}
\newcommand{\mchat}{\ensuremath{{\hat m}_c}}
\newcommand{\muhat}{\ensuremath{{\hat m}_u}}
\newcommand{\mdhat}{\ensuremath{{\hat m}_d}}

\newcommand{\mbhat}{\ensuremath{{\hat m}_b}}
\newcommand{\mqhat}{\ensuremath{{\hat m}_q}}
\newcommand{\mlhat}{\ensuremath{{\hat m}_l}}

\newcommand{\gc}{\ensuremath{g(\mchat,\shat)}}
\newcommand{\gu}{\ensuremath{g(\muhat,\shat)}}
\newcommand{\gd}{\ensuremath{g(\mdhat,\shat)}}

\newcommand{\gb}{\ensuremath{g(\mbhat,\shat)}}
\newcommand{\gq}{\ensuremath{g(\mqhat,\shat)}}

\newcommand{\yq}{\ensuremath{{y_{q}}}}
\def\msbar{\ensuremath{{\rm{\overline{MS}}}}} 
\newcommand{\Imag}{\ensuremath{{\operatorname{Im}}}}
\newcommand{\Real}{\ensuremath{{\operatorname{Re}}}}
\newcommand{\BABAR}{\ensuremath{\text{BaBar}\ }}
\newcommand{\BELLE}{\ensuremath{\text{Belle}\ }}
\newcommand{\Cixeff}{\ensuremath{C_9^{\rm eff}}}
\newcommand{\Cviieff}{\ensuremath{C_7^{\rm eff}}}

\newcommand{\lsim}{{\;\raise0.3ex\hbox{$<$\kern-0.75em\raise-1.1ex\hbox{$\sim$}} 
\;}} 
\newcommand{\gsim}{{\;\raise0.3ex\hbox{$>$\kern-0.75em\raise-1.1ex\hbox{$\sim$}} 
\;}} 
 
\begin{document} 
\thispagestyle{empty}
\rightline{hep-ph/0304077}
\rightline{DO-TH 02/21}
\rightline{OSU-HEP-03-03}
\rightline{DESY 03-046}
\rightline{April 2003}
\vspace*{0.5cm}

\begin{center}
{\large \bf Standard Model $CP$ violation in Polarised 
$b\rightarrow d l^+ l^-$}

\medskip

\end{center}
\medskip
\begin{center}
{K.S. Babu\footnote{babu@hep.phy.okstate.edu}}\\
{\it Deptartment of Physics, Oklahoma State University, 
Stillwater, OK 74078, U.S.A.}\\
{K.R.S. Balaji\footnote{balaji@zylon.physik.uni-dortmund.de},}\\
{\it  Institut f\"ur Theoretische Physik, Universit\"at Dortmund,
44221 Dortmund,\\ Germany}\\ and\\
\smallskip
{I. Schienbein\footnote{schien@mail.desy.de}}\\
{\it DESY/Univ.\ Hamburg, Notkestrasse 85, 22603 Hamburg, Germany}\\
\end{center}
\noindent
\begin{abstract} 
In the standard model, we study $CP$ violating rate asymmetries 
in the decay $b \to d l^+ l^-$ when one of the  leptons is polarised. 
We find an asymmetry of 
$(5\div 15)\%$ in the polarised decay spectrum which is comparable to known
results for the unpolarised case. In the kinematic region separating the
$\rho-\omega$ and $c \bar c$ resonances, which is also theoretically cleanest,
the polarised contribution
to the asymmetry is larger than the unpolarised results. 
In order to observe a $3\sigma$ signal for direct CP violation in the
polarised spectrum, assuming $100\%$ efficiency, about $10^{10}$ $B\bar B$ 
pairs are required at a B factory. Our results
indicate an asymmetric contribution from the individual polarisation states
to the unpolarised $CP$ asymmetry. Taking advantage of this, one can 
attribute any new physics to be most sensitive to a specific polarisation
state. 
\end{abstract}


\newpage 
 
\section{Introduction} 

Flavor changing neutral current decays are important probes for new physics 
beyond 
the standard model (SM) \cite{Buras:2001pn}. 
%
Such currents are absent in the SM at 
the tree level and their strength is small at the one-loop level
due to the GIM mechanism \cite{Glashow:1970gm}.
Furthermore, such transitions 
can also be CKM suppressed in the SM. Hence, in such decays, any enhancement 
over the SM expectations is an unambiguous signature for new physics. In this 
direction, the first observations of the radiative B decays 
$( B \rightarrow X_s \gamma)$ by CLEO \cite{Alam:1995aw} and later by 
ALEPH \cite{Barate:1998vz} have yielded $|V_{ts}V_{tb}^*| \sim 0.035$ which 
is in conformity with the CKM estimates.
However, in radiative decay modes, deviations from SM predictions are 
sensitive to the photon spectrum which posses considerable experimental
difficulties, in particular, measuring the lower end of the 
spectrum \cite{Neubert:1998zx}.
In parallel, rare semi-leptonic decays $(b \rightarrow q l^+l^-)$ can provide 
alternative sources to discover new physics and these are relatively cleaner
than pure hadronic decays. 
The semi-leptonic
decays are induced by strong interaction corrected effective Hamiltonians 
and the matrix elements receive the dominant contributions from 
loops with virtual top quarks and weak bosons
which are proportional to $V_{tb}V_{tq}^{*}$ with $q = d, s$. 
 In such decays, the standard  
observables such as lepton polarization asymmetries, forward-backward 
asymmetry and also $CP$ violation can be studied kinematically as a function
of the invariant di-lepton mass.  
Within the SM, the semi-leptonic process has earlier been studied by a number 
of authors \cite{Deshpande:1989bd,*Ali:1991fy,*Greub:1995pi,*Melikhov:1997zu,*Roberts:1996ie,*Geng:1996az,*Burdman:1998mk}. 

In the SM, for 
the decay $B \to X_{q} l^{+} l^{-}$, the dominant
contribution is from the parton level process $b \to q l^+ l^-$ 
and is defined in terms of a set of ten effective operators
$O_1 - O_{10}$ \cite{Witten:1977kx}. To be specific, we shall take $q=d$, and
the truncated operator basis is reduced to the following five operators
\cite{Grinstein:1989me} given to be

\begin{eqnarray}
O_1 ~=~ ( \bar d^ {\alpha}_L \gamma_{\mu} b^ {\alpha}_L) ( \bar c^
        {\beta}_L\gamma^{\mu} c^ {\beta}_L)~,
\nonumber\\ 
O_2 ~=~ ( \bar d^
        {\alpha}_L \gamma_{\mu} b^ {\beta}_L) ( \bar c^ {\alpha}_L
        \gamma^{\mu} c^ {\beta}_L)~,\nonumber\\ O_7 ~=~ ( \bar d^{\alpha}
        \sigma_{\mu\nu}[m_b P_R + m_d P_L] b^ {\alpha}) F^{\mu\nu}~,
\nonumber\\
O_9 ~=~ ( \bar d^{\alpha}_L \gamma_{\mu} b^{\alpha}_L)(\bar
        l\gamma^{\mu} l)~,
\nonumber\\ 
O_{10}~=~ ( \bar d^{\alpha}_L
        \gamma_{\mu} b^{\alpha}_L)(\bar l \gamma^{\mu} \gamma_{5}l)~,
\label{opdef1}
\end{eqnarray}
and the additional operators \cite{Kruger:1997dt}
\begin{eqnarray}
O_1^u ~=~ 
( \bar d^{\alpha}_L \gamma_{\mu} u^ {\beta}_L) 
( \bar u^{\beta}_L \gamma^{\mu} b^{\alpha}_L)~,
\nonumber\\
O_2^u ~=~ 
( \bar d^{\alpha}_L \gamma_{\mu} u^{\alpha}_L) 
( \bar u^{\beta}_L \gamma^{\mu} b^{\beta}_L)~.
\end{eqnarray}
In the presence of new physics, there can be new tree level operators due 
to possible right-handed current-current operators and also operators 
involving SUSY scalar particles \cite{dafne}. In this analysis, we do not
consider them and will limit to the interactions determined by the SM. 

Prior to the advent of the B factories such as \BABAR and \BELLE 
the best experimental limits 
for the inclusive branching ratios $BR(b \to s l^+ l^-)$ 
with $l = e,\mu$ as measured by CLEO \cite{Glenn:1998gh} 
have been an order higher than the SM estimates \cite{Ali:1997bm}. 
Recently, the first measurement of this decay has been reported 
by \BELLE \cite{Kaneko:2002mr} and is in agreement with the SM expectations
and hence further constrains any extensions to the SM. In this context, it is
pertinent to exploit the features of the SM which makes it most sensitive to
new physics. It has been argued that measuring the various lepton polarizations
provides for a comprehensive test of the SM and further can be indicative 
of new physics \cite{Bensalam:2002ni}.

 In this exercise, we examine the SM contribution to $CP$ violating 
effects for the case when 
one of the leptons is observed in a polarised state. As it turns out, the 
CP asymmetry is comparable to the unpolarised SM expectations when the lepton 
is in a specific polarised state. In other words, the net unpolarised 
$CP$ asymmetry (which is a linear sum of the polarised asymmetries) receives
unequal contributions from the individual polarised state. This feature can   
provide for measurements involving new physics search.
Note that this conclusion, which is dependent 
on a specific polarisation state of the lepton is apriori not obvious from 
the operator structure given in (\ref{opdef1}); albeit, is well motivated from 
the observation that the electroweak sector of the SM is 
polarised (left-handed).
Therefore, measurements of decay spectra when one of the final 
state lepton is in the 
{\it wrong sign } polarised state will be a key observable to any new physics. 
By {\it wrong sign} polarised state, we mean a lepton with a given polarisation,
whose SM decay width is smaller when compared to the unpolarised spectrum.

 Our paper is organized as follows. In the next section, we revise the 
basic theoretical framework and describe various observables such as
decay spectra for polarised and unpolarised leptons and polarisation 
asymmetries. In order to be self contained, we have updated the
known results by recalculating all of the existing results taking care of 
the various theoretical issues involved. In section \ref{sec:CP}, we derive the
CP violating decay rate asymmetries for polarised final state leptons
and compare this with the CP violating asymmetry in the unpolarised case.
In section \ref{sec:results}, we present our numerical results 
and estimate the feasibility of measuring the $CP$ violation in a B factory. 
Finally, in section \ref{sec:summary}, we conclude with a summary of the main 
results.

\section{Theoretical framework}\label{sec:framework}
In this section, we describe our procedure for 
obtaining the decay spectra. We first  perform a detailed analysis and
rederive the known parton level results. In doing so, we try to clarify 
some of the theoretical uncertainties which go as input parameters for 
our numerical estimates. Our purpose is to ensure that our conclusions have 
subsumed some of these theoretical difficulties and also to complement the
existing literature.\\

The QCD corrected effective Hamiltonian describing the decay
$b \to d l^+ l^-$  in the SM 
leads to the matrix element \cite{Kruger:1997dt,Buras:1995dj}
\begin{eqnarray}
M &=&  \frac{G_{F} \alpha V_{tb}V_{td}^*}{\sqrt{2} \pi}
\Big[{\Cixeff}(\bar{d} \gamma_\mu P_{L} b) \bar{l} \gamma^{\mu} l  
    +C_{10}(\bar{d} \gamma_\mu P_{L} b)  \bar{l} \gamma^{\mu} \gamma^{5}l 
\nonumber\\*
& &\phantom{=  \frac{G_{F} \alpha V_{tb}V_{tq}^*}{\sqrt{2} \pi}\Big[}
- 2 {\Cviieff} \bar{d} i\sigma_{\mu\nu} \frac{q^{\nu}}{q^2} (m_{b}P_{R} + 
  m_{d}P_{L})b \bar{l} \gamma^{\mu} l\Big]~,
\label{eq:heff}
\end{eqnarray}
where the notations employed are the standard ones and $q$ denotes the 
four momentum of the lepton pair.
In the SM, except for $\Cixeff$, the Wilson couplings are real and analytic 
expressions can be found in the literature 
\cite{Buras:1995dj,Buchalla:1996vs,misiak}. 
The numerical values of the Wilson coefficients depend on five parameters
($\mu$, $m_t$, $m_W$, $\sin^2 \theta_W$, $\alpha_s(M_Z)$)
which are listed in the appendix of this paper.
For a consistent NLO analysis all the coefficients, except $C_9$, should be
calculated in the leading log approximation; see \cite{Buras:1995dj}
for a detailed explanation. We obtain these to be,
\be
\begin{gathered}
\Cviieff = - 0.310\ ,\quad C_{10} = -4.181\ ,\quad
C_1 = -0.251\ ,\quad  C_2 = 1.108
\\ 
C_3 = 1.120 \times 10^{-2}\ ,\quad
C_4 = -2.584 \times 10^{-2}\ ,\quad C_5 = 7.443 \times 10^{-3}\ ,
\\
C_6 = -3.167 \times 10^{-2}\ ,\quad C_9^{\rm NDR} = 4.128 \ .
\label{eq:coeff}
\end{gathered}
\ee
We place a few remarks here. The above numbers differ from 
the ones used in \cite{Kruger:1997dt} and we attribute the main source of 
this difference in the use of the $\msbar$ top quark mass 
$m_t = 165\ \gev$ \cite{Buras:1995dj} instead of a pole mass of 
$m_t = 176\ \gev$. It is noteworthy that 
the leading log results have been obtained
utilizing a two-loop ($\msbar$) $\alpha_s$ with $\alpha_s(M_Z^2)=0.1183$.
Using a leading order $\alpha_s$ 
with $\alpha_s^{\rm LO}(M_Z^2) = 0.130$ as obtained by the
CTEQ collaboration from a global LO analysis of parton distribution 
functions \protect\cite{Pumplin:2002vw} results in marginal changes, e.g., 
$C_1 =-0.261$, $C_2 = 1.114$, $\Cviieff = -0.314$.
The value of $C_9 \equiv C_9^{\rm NDR}$ has been obtained at NLO 
accuracy in the NDR scheme. 
On the other hand, the effective coefficient $\Cixeff$ is scheme independent.
It can be parametrized in the following way \cite{Kruger:1997dt}:
\be
\Cixeff = \xi_1 + \lu \xi_2\ , \quad \lu = 
\frac{V_{ub}V_{ud}^*}{V_{tb}V_{td}^*}~,
\label{eq:c9eff}
\ee
with 
\ba
\xi_1 & = & C_9  + 0.138\ \omega(\shat) + \gc (3 C_1 + C_2 + 3 C_3 + C_4 + 3 C_5 + C_6)
\nonumber\\
  && - \frac{1}{2} \gd (C_3 + C_4) - \frac{1}{2} \gb (4 C_3 + 4 C_4 + 3C_5 + C_6)
\label{eq:xi1}\\
  && + \frac{2}{9} (3 C_3 + C_4 + 3C_5 + C_6)\ ,
\nonumber
\\
\xi_2 & = & [\gc - \gu](3 C_1 + C_2)\ .
\label{eq:xi2}
\ea
Here, $\shat = q^2/m_b^2$ and $\mqhat = m_q/m_b$ are dimensionless variables 
scaled with respect to the bottom quark mass. Note that
some of the Wilson coefficients in \protect\eqref{eq:coeff} entering 
 \protect\eqref{eq:xi1} and \protect\eqref{eq:xi2} 
are small and/or add up destructively in 
such a way that the functions $\xi_{1,2}$ in the NLO
approximation give the following simple expressions,
\be
\xi_1 \simeq 4.128 + 0.138\ \omega(\shat) 
+ 0.36\ g(\mchat, \shat),~
\xi_2 \simeq 0.36\ [g(\mchat, \shat) - g(\muhat, \shat)]~.
\label{approx}
\ee

In \eqref{eq:xi1}, the function $\omega(\shat)$ represents one loop corrections
to the operator $O_9$ \cite{jezabek:1989ja} and the function $g(\mqhat, \shat)$
represents the corrections to the four-quark operators $O_1-O_6$ \cite{misiak} , i.e.,
\ba
\gq &=& -\frac{8}{9} \ln(\mqhat) + \frac{8}{27} + \frac{4}{9}\ \yq
          - \frac{2}{9} (2 + \yq) \sqrt{|1-\yq|}\
\bigg\{\Theta(1-\yq) 
\times
\nonumber\\
&&\left[ \ln\left(\tfrac{1+\sqrt{1-\yq}}{1-\sqrt{1-\yq}}\right) 
- i \pi \right]
 + \Theta(\yq - 1)\ 2 \arctan{\frac{1}{\sqrt{\yq-1}}}
\bigg\}~,
\label{eq:gcont_ana}
\ea
with $\yq \equiv 4 \mqhat^2/\shat$.


In addition to the short distance contributions described above,
the decays $B \to X_{d} l^{+} l^{-}$
also receive long distance
contributions from the tree-level diagrams involving 
$u \bar u$, $d \bar d$, and $c \bar c$
bound states, 
$B \to X_d  (\rho, \omega, J/ \psi, \psi^{\prime}, ...) \to  X_d l^+l^-$. 
These effects can be taken into account by modifying
the functions $\gq$.
In the case of the $J/\Psi$ family
these pole contributions can be
incorporated by employing a Breit-Wigner form for the resonance states
\cite{Deshpande:1989bd,*Ali:1991fy,*Greub:1995pi,*Melikhov:1997zu,*Roberts:1996ie,*Geng:1996az,*Burdman:1998mk,Lim:1989yu} through the replacement
\begin{equation}
\gc \rightarrow \gc - 
\frac{3 \pi}{\alpha ^2} 
\sum_{V=J/\psi, \psi^{\prime}, ...}  
\frac{ M_V Br(V \rightarrow l^+l^-)
\Gamma^V_{all}}{s - M_V^2 ~+~ i \Gamma^V_{all}M_V}~.
\label{BWC}
\end{equation}
In (\ref{BWC}), the sum includes all $c \bar c$ bound states $V$ with mass 
$M_V$ and total width $\Gamma^V_{all}$.
The $c \bar c$ long distance contributions to the branching fraction
are obviously dominant near the $J/ \psi$ resonance peak. In addition
to this, there can be long distance contributions from real $u \bar
u$, $ d \bar d$ quark pairs giving rise to intermediate vector meson
states, i.e., $\rho$ and $\omega$. In many of the existing analysis
performed so far, these resonance contributions have been neglected.
A plausible reason being, in the decays $B \to X_s l^+ l^-$ they are CKM 
suppressed. However, in \cite{Kruger:1997dt} it has been demonstrated that 
the $\rho$ and $\omega$ resonances contribute quite sizeably to
the CP violating decay rate asymmetry and should be taken
into account in theoretical studies of this observable.

Following the approach in \cite{Kruger:1997dt,Kruger:1996cv}
we calculate the functions $\gu$, $\gd$ and $\gc$ using
a dispersive method. To be precise, $\gu$ and $\gd$ have been calculated 
exactly along the lines described in \cite{Kruger:1997dt,Kruger:1996cv} 
using quark masses $m_u = m_d = m_\pi$.
Concerning $\gc$, the continuous part has been computed 
with the dispersive method as well whereas, for simplicity, 
the $c \bar c$ resonances have been evaluated according to \eqref{BWC}.
Finally, the function $\gb$ has been evaluated
with help of \eqref{eq:gcont_ana}. Note that the contribution
of $\gb$ to $\Cixeff$ in \eqref{eq:c9eff} is negligible.


\subsection{Unpolarized decay spectrum}
In the absence of low energy QCD
corrections ($\sim 1/m_b^2$) \cite{Falk:1994dh,Ali:1997bm} and 
setting the down quark
mass to zero, the unpolarized differential decay width as a function of the
invariant mass of the lepton pair is
\begin{equation}
\frac{d\Gamma}{d{\shat}} =  \frac{G_{F}^2 m_{b}^5 \alpha^2}{768
\pi^5}|V_{tb}V_{td}^{*}|^2 \lambda^{\frac{1}{2}}(1, {\shat},0)\
a\
\Delta(\shat)~;\quad
a \equiv \sqrt{1 - \frac{4 {\mlhat}^2}{\shat}}~. 
\label{diffspec}
\end{equation}  
In (\ref{diffspec}), the triangle function $\lambda$ is given 
$\lambda(a,b,c)= a^2 + b^2+ c^2 -2(ab+bc+ca)$ and the kinematic factors are 
\begin{eqnarray}
\Delta(\shat) &=& \big[12\Real({\Cviieff} {\Cixeff}^*)F_1(\shat) +
\frac{4}{\shat}|{\Cviieff}|^2
F_2(\shat)\big] (1+\frac{2 \mlhat^2}{\shat})\nonumber\\
& &+(|{\Cixeff}|^2 +|C_{10}|^2)F_3(\shat) 
+6\mlhat^2(|{\Cixeff}|^2 -|C_{10}|^2)F_1(\shat)~,\nonumber\\ 
F_1(\shat) &=&  1 -\shat~,\nonumber\\
F_2(\shat) &=&  2 - \shat - \shat^2~,\nonumber\\ 
F_3(\shat) &=&  1
+\shat -2\shat^2  + \lambda(1, \shat,0)
\frac{2\mlhat^2}{\shat}~.
\label{eq:defFunpoll}
\end{eqnarray}
The various quantities in the above differential decay width are
scaled with respect to the $b$ quark mass $m_b$ and are indicated by a
hat. 
The physical range for $\shat$ is given by $4\mlhat^2 \le \shat \le 1$
as is dictated by the threshold factor, $a$, and the triangle function in
(\ref{diffspec}).
The above differential spectrum has been obtained earlier by
Kr\"uger et al., \cite{Kruger:1996cv,Kruger:1997dt} (for $m_l \neq 0$ and 
$m_q \neq 0$),
Ali et al., \cite{Ali:1991is,Ali:1995bf} (for $m_l=0$), Grinstein et al.,
\cite{Grinstein:1989me}, Buras et al., \cite{Buras:1995dj} (for $m_q=0$ and 
$m_l=0$) 
and by Hewett \cite{Hewett:1996dk} (for $m_q =0$, which agrees with
our expressions). 

As usual we remove uncertainties in \eqref{diffspec} due to the
$b$ quark mass (a factor of $m_b^5$) by introducing the charged current 
semi-leptonic decay rate
\be
\Gamma(B \to X_c e \bar \nu_e)= \frac{G_{F}^2 m_{b}^5}{192 \pi^3}
|V_{cb}|^2 f(\mchat) \kappa(\mchat)
\ee
where $f(\mchat)$ and $\kappa(\mchat)$ represent the phase space
and the one-loop QCD corrections to the semi-leptonic decay and
can be found in \cite{Kruger:1997dt}.
Therefore the differential branching ratio can be written as
\ba
\frac{d \Gamma}{d \shat} =  \frac{\alpha^2}{4 \pi^2}
\frac{|V_{tb}V_{td}^{*}|^2}{|V_{cb}|^2} 
\frac{B(B \to X_c e \bar \nu_e)}{f(\mchat)\kappa(\mchat)}
 \lambda^{\frac{1}{2}}(1, {\shat},0)\ a\ \Delta(\shat)\ .
\label{eq:BR}
\ea

\subsection{Polarized decay spectrum}
The lepton polarisation has been first analyzed by
Hewett \cite{Hewett:1996dk} and Kr\"uger and Sehgal \cite{Kruger:1996cv}
who
showed that additional information 
can be obtained on the quadratic functions of the effective Wilson couplings,
${\Cviieff}$, $C_{10}$ and ${\Cixeff}$. 
In order to calculate the polarised decay spectrum, one defines a reference
frame with three orthogonal unit vectors $\eL$, $\eN$ and $\eT$, such that
\begin{eqnarray}\label{eq:unitvecdef}
\eL &=& \frac{\pl}{|\pl|}~,
\nonumber\\
\eN &=&\frac{\pq \times \pl}{|\pq \times \pl|}~,
\\
\eT &=& \eN \times \eL ~,
\nonumber
\end{eqnarray}
where $\pq$ and $\pl$ are the three momentum vectors of the
quark and the $l^-$ lepton, respectively, in the $l^+ l^-$ center-of-mass
system.
Given the lepton $l^-$
spin direction $\vecn$, which is a unit vector in the $l^-$ rest
frame, the differential decay spectrum can be written as
\cite{Kruger:1996cv}
\begin{equation}
\frac{d\Gamma(\shat, \vecn)}{d\shat} = \frac{1}{2}
\left(\frac{d\Gamma(\shat)}{d\shat}\right)_{\rm unpol} 
\Big[ 1 + (\PL \eL + \PT \eT + \PN \eN) \cdot \vecn \Big]~,
\label{eq:poldecay}
\end{equation}
where
the polarisation components $\PL$, $\PN$ and $\PT$ can be
constructed as follows;
\begin{equation}
\PI(\shat) =  \frac{d \Gamma (\vecn = \ei)/d \shat  -  d
\Gamma (\vecn = -\ei)/d\shat} {d \Gamma (\vecn = \ei)/d\shat  
+  d \Gamma (\vecn = -\ei)/d\shat}~, {\rm ~i = L, ~N, ~ T~}.
\label{leppolldef}
\end{equation}

The resulting polarisation asymmetries are
\begin{eqnarray}
\PL(\shat)&=& \frac{a}{\Delta(\shat)} 
\Big[12\Real({\Cviieff} C_{10}^*)(1-  \shat ) +  2\Real({\Cixeff}C_{10}^*)(1 ~ + 
\shat -2 \shat^2)\Big],
\nonumber \\ 
\PT(\shat)  &=& \frac{3 \pi \mlhat}{2 \Delta (\shat)
\sqrt{\shat}} \lambda^{1/2}(1, \shat,0) \Big[2\Real({\Cviieff} C_{10}^*) -4
\Real({\Cviieff} {\Cixeff}^*) - \frac{4}{\shat}|{\Cviieff}|^2 
\nonumber\\
&\phantom{=}&+ \Real({\Cixeff}C_{10}^*)- |{\Cixeff}|^2 \shat\Big],
\nonumber\\ 
\PN(\shat)  &=&  \frac{3 \pi \mlhat a}{2 \Delta (\shat)}
\ \lambda^{1/2}(1, \shat,0)\ \sqrt{\shat}\ \Imag({\Cixeff}^*C_{10})~,
\label{leppollasym}
\end{eqnarray}
where we differ by a factor of 2 in $\PT$ with respect to the
results obtained in \cite{Kruger:1996cv}. The above expressions for $\PI$ 
agree with \cite{dafne} for the SM case. 

As can be seen, the polarisation asymmetries
in (\ref{leppollasym}) have different quadratic
combinations of the Wilson couplings and any alteration in the values
of these couplings can lead to changes in the asymmetries. Thus, these
are sensitive to new physics and can also probe the relative signs of
the couplings ${\Cviieff}$, ${\Cixeff}$ and $C_{10}$. The normal polarisation
asymmetry $\PN$ is proportional to $\Imag({\Cixeff}C_{10}^*)$ and is thus
sensitive to the absorptive part of the loop contributed by the charm
quark. 

The transverse and normal asymmetries $\PT$ and $\PN$, respectively,  are 
proportional to
$\mlhat$ and thus the effects can be significant for the case of
tau leptons. However, the case for tau leptons is an experimental
challenge. One reason is that the tau decays in to final states with a 
neutrino and this involves uncertainties due to missing energy. Also,
the predominant decays of the tau lepton are hadronic and this is an
undesirable feature at hadronic colliders. Setting aside these problems, in 
our analysis, we illustrate the results when the final state lepton is an 
electron and tau; the results when a muon is produced are almost
identical to the electron case.


\section{CP violation}\label{sec:CP}
In the SM, $CP$ violation in the decay $B \to X_{s} l^{+} l^{-}$ is 
strongly suppressed. This follows
from the unitarity of the $CKM$ matrix. However, in the semi-leptonic
$B$ decay, $B \to X_{d} l^{+} l^{-}$, the $CP$
violating effects are not so strongly suppressed. 
The $CP$ asymmetry for this decay can be observed by measuring the partial 
decay rates for the
process and its charge conjugated process \cite{Kruger:1997dt,Ali:1998sf}. 
Before turning to a derivation of CP violating asymmetries 
for the case of polarised final state leptons 
it is helpful to recall the unpolarised case.
\subsection{Unpolarized decay spectrum}
In the unpolarised case the CP-violating rate asymmetry can be defined by
\begin{equation}
\ACP = \frac{\Gamma_0 - \bar\Gamma_0}{\Gamma_0 + \bar\Gamma_0}~,
\end{equation}
where
\begin{equation}
\Gamma_0 \equiv \frac{d\Gamma}{d \shat} \equiv
\frac{d\Gamma(b\to d l^+ l^-)}{d \shat}\ , \
\bar \Gamma_0 \equiv \frac{d\bar\Gamma}{d \shat} \equiv
\frac{d\Gamma({\bar b} \to {\bar d} l^+ l^-)}{d \shat}~.
\label{eq:gamma0}
\end{equation}
The explicit expression for the unpolarized particle decay rate 
$\Gamma_0$ has been given in (\ref{diffspec}).
Obviously, it can be written as a product of a real-valued
function $r(\shat)$ times the 
function $\Delta(\shat)$, given in (\ref{eq:defFunpoll});
$\Gamma_0(\shat) = r(\shat) \ \Delta(\shat)$. Taking the approach as 
in \cite{Kruger:1997dt} we write the matrix elements for the decay and 
the anti-particle decay as
\begin{equation}
M = A + \lambda_u B\ , \ \bar M = A + \lambda_u^* B~,
\label{eq:CPMdef2}
\end{equation}
where the CP-violating parameter
$\lu$, entering the Wilson coupling ${\Cixeff}$, has been 
defined in \eqref{eq:c9eff}.
Consequently, the rate for the anti-particle decay is then given by
\be
\bar \Gamma_0 = {\Gamma_0}_{|\lu \to \lu^*}
= r(\shat) \bar \Delta(\shat)~;
~\bar\Delta=\Delta_{|{\lambda_u \to \lambda_u^*}}~.
\label{apr}
\ee
 
Using (\ref{diffspec}) and (\ref{apr}), the CP violating asymmetry is evaluated
to be \cite{Kruger:1997dt} 
\begin{equation}
\ACP(\shat) =  \frac{\Delta - \bar \Delta}{\Delta + \bar \Delta}
= \frac{-2 \Imag( \lu) \Sigma}{\Delta + 2 \Imag( \lu) \Sigma}
\simeq -2 \Imag( \lu) \frac{\Sigma(\hat s)}{\Delta (\hat s)}~.
\label{eq:CPdef}
\end{equation}
In (\ref{eq:CPdef}), 
\begin{equation}
\Sigma(\shat) = \Imag(\xi_1^* \xi_2)[F_3(\shat) + 
6 \mlhat^2 F_1(\shat)]
+6\Imag({\Cviieff} \xi_2)F_1({\shat})(1+ \frac{2 \mlhat^2}{\shat})~.
\label{eq:CPde1}
\end{equation}
In the presence of lepton $l^-$ polarisation, the above 
$CP$ asymmetry can get modified and receives a contribution from
$C_{10}$ through the interference piece with ${\Cixeff}$ 
in $|M|^2$; see~(\ref{eq:poldecay}) and (\ref{leppollasym}). 
In the following, we discuss this modification.

\subsection{Polarized decay spectrum}
In the polarised case, a CP violating asymmetry can be defined as
follows
\begin{equation}
\ACP(\vecn) = \frac{\Gamma(\vecn) - \bar\Gamma(\bar \vecn = - \vecn)}
{\Gamma_0 + \bar\Gamma_0}~,
\label{eq:acpn}
\end{equation}
where
\begin{eqnarray}
\Gamma(\vecn) &\equiv& \frac{d\Gamma(\shat,\vecn)}{d \shat}
\equiv \frac{d\Gamma(b \to d l^+ l^-(\vecn))}{d \shat}\ ,
\nonumber\\
\bar\Gamma(\bar \vecn) &\equiv& \frac{d\bar\Gamma(\shat,\bar\vecn)}{d \shat}
\equiv \frac{d\Gamma(\bar b \to \bar d l^+(\bar\vecn) l^-)}{d \shat}\ ,
\end{eqnarray}
and $\Gamma_0$, $\bar\Gamma_0$ have been defined in the previous section.
Further, $\vecn$ is the spin direction of the lepton $l^-$ in the
$b$-decay and $\bar \vecn$ is the spin direction of the $l^+$ in the
$\bar b$-decay.
For instance, assuming CP conservation, the rate for the decay
of a $B$ to a left handed electron should be the same as the rate
for the decay of a $\bar B$ to a right handed positron.
More generally, in the CP conserving case, we would have
\begin{equation}
\Gamma(\vecn) \overset{CP\ cons.}{=} \bar \Gamma(\bar \vecn = -\vecn)\ 
\Rightarrow\ \ACP(\vecn) = 0 \ .
\end{equation}
The polarised decay spectrum for the $b$-decay has already been stated
in (\ref{eq:poldecay}).
Using the above definitions the spectrum reads,
\begin{equation}
\Gamma(\vecn) = \tfrac{1}{2} \Gamma_0\ (1 + \PI\ \ei \cdot \vecn)~,
\label{eq:gamma1}
\end{equation}
where a sum over $\rm i = L, T, N$ is implied. 
Analogously, for the corresponding $CP$ conjugated process we have the 
decay spectrum 
\begin{equation}
\bar\Gamma(\bar\vecn) = \tfrac{1}{2} \bar\Gamma_0\ 
(1 +\bar \PI\ \eibar \cdot \bar\vecn)\ .
\label{eq:gamma2}
\end{equation}
Note that (\ref{eq:gamma1}) and (\ref{eq:gamma2}) {\em define} the
polarisation asymmetries $\PI$ and $\bar \PI$, respectively.
With the choice $\eibar = \ei$, the $\bar \PI$ are 
constructed from the decay spectrum in
complete analogy to the $\PI$ in (\ref{leppolldef}).
Moreover, from the condition 
$\Gamma(\vecn) = \bar \Gamma(\bar \vecn = -\vecn)$, in the CP
conserving case, it follows that $\bar \PI = - \PI$.
In the general case we have
\begin{equation}
\bar \PI = - {\PI}_{|\lu \to \lu^*} \quad 
(\text{for} ~\eibar = \ei)\ .
\end{equation} 

Now everything is at hand to calculate the CP violating
asymmetries for a lepton $l^-$ with polarisation $\vecn = \ei$;
inserting (\ref{eq:gamma1}) and (\ref{eq:gamma2}) into
(\ref{eq:acpn}) one obtains the asymmetry,
\begin{eqnarray}
\ACP(\vecn =\pm \ei) &=& 
\frac{1}{2} 
\left[\frac{\Gamma_0 - \bar\Gamma_0}{\Gamma_0 + \bar \Gamma_0} 
\pm \frac{\Gamma_0 \PI - (\Gamma_0 \PI)_{|\lu \to \lu^*}}
{\Gamma_0 + \bar \Gamma_0} \right]
\nonumber\\
&=&\frac{1}{2} 
\left[\ACP(\hat s) \pm \frac{\Delta \PI - (\Delta \PI)_{|\lu \to \lu^*}}
{\Delta(\hat s) + \bar \Delta (\hat s)} \right]
\nonumber\\
&\equiv&
\frac{1}{2} \left[ \ACP(\hat s) \pm \delta\ACP^{\rm i} (\hat s)\right] \ , 
\quad {\rm i = L,T,N}~.
\label{eq:acpei}
\end{eqnarray}
On the r.h.s.\ of (\ref{eq:acpei}), $A_{CP}(\hat s)$ is the unpolarized
CP violating asymmetry given in (\ref{eq:CPdef}).
The polarised quantities $\delta\ACP^{\rm i}(\hat s)$ denote the modifications
to the unpolarised spectra and will be stated explicitly below.
It should be noted that for a given polarisation there are two 
independent observables, $\ACP(\vecn =\ei)$ and $\ACP(\vecn =- \ei)$
or, alternatively, $A_{CP}(\hat s) = \ACP(\vecn =\ei)+\ACP(\vecn =- \ei)$ 
(as it must be) and 
$\delta\ACP^{\rm i}(\hat s) =  \ACP(\vecn =\ei)-\ACP(\vecn =- \ei)$.
The polarised CP violating asymmetries can be evaluated
by inspecting the polarisation asymmetries in (\ref{leppollasym}).
After some algebra one finds the following final results:
\begin{equation}
\delta\ACP^{\rm i} (\hat s)
= \frac{-2 \Imag( \lu) \delta\Sigma^{\rm i}(\hat s)}
{\Delta (\hat s) + 2 \Imag( \lu) \Sigma(\hat s)}
\simeq -2 \Imag( \lu) \frac{\delta\Sigma^{\rm i}(\hat s)}{\Delta (\hat s)}~,
\label{eq:polCPdef}
\end{equation}
with
\begin{eqnarray}
\delta\Sigma^{\rm L}(\hat s) & = &
\Imag(C_{10}\xi_2)\ (1+\shat -2 \shat^2) \ a\ ,
\nonumber\\
\delta\Sigma^{\rm T} (\hat s)& = &
\frac{3 \pi \mlhat}{2 \sqrt{\shat}} \lambda^{1/2}(1,\shat, 0)
\left[-2 \Imag({\Cviieff} \xi_2) + \frac{1}{2} \Imag(C_{10}\xi_2)-
\shat \Imag(\xi_1^* \xi_2)\right]\ ,
\nonumber\\
\delta\Sigma^{\rm N} (\hat s)& = &
\frac{3 \pi \mlhat}{2 \sqrt{\shat}} \lambda^{1/2}(1,\shat, 0)
\left[\frac{\shat}{2} \Real(C_{10} \xi_2)\right] a\ ,
\label{eq:sigmai}
\end{eqnarray}
where $a$ is the threshold factor defined in (\ref{diffspec}). It is
interesting to note that the asymmetries, $\delta\Sigma^{\rm T}(\hat s)$ and 
$\delta\Sigma^{\rm N}(\hat s)$, have different combinations involving the 
imaginary and real parts of $\xi_2$; thereby, show dependence on the 
corrections to the four-quark operators, $O_1 - O_6$.   

\section{Numerical analysis and discussion}
\label{sec:results}
\renewcommand{\arraystretch}{1.2}

\begin{figure}[ht]
\begin{center}
\hspace*{-2.0cm}
\epsfig{file=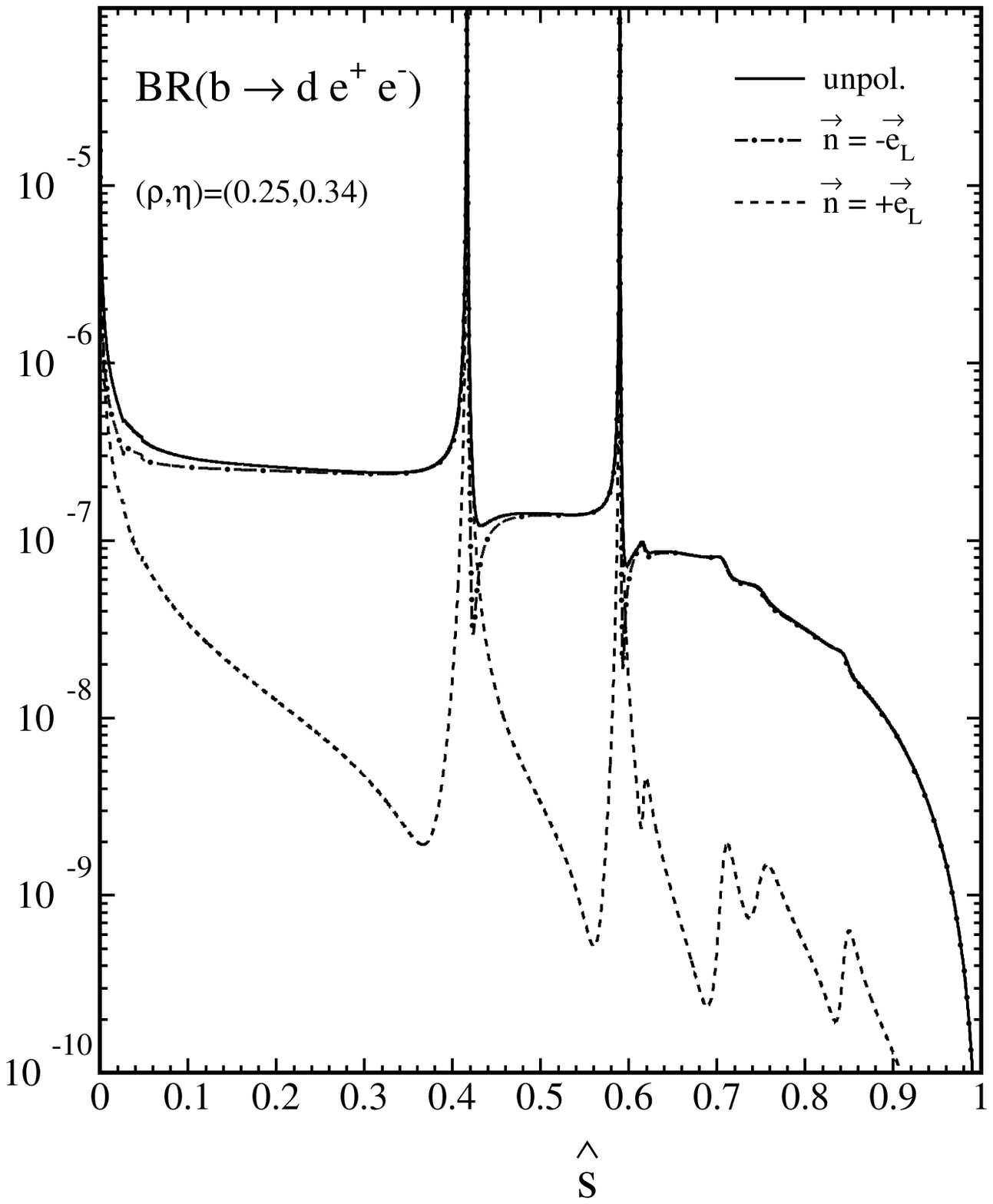,angle=0,width=12cm}
\end{center}
\vspace*{-0.5cm}
\caption{Polarised and unpolarised branching
ratios for the decay $b \to d e^+ e^-$
according to \protect\eqref{eq:BR}
and \protect\eqref{eq:poldecay}.
The unit vector $\eL$ has been defined in 
\protect\eqref{eq:unitvecdef}.}
\label{fig:br_e} 
\end{figure} 

\begin{figure}[ht]
\begin{center}
\hspace*{-2.0cm}
\epsfig{file=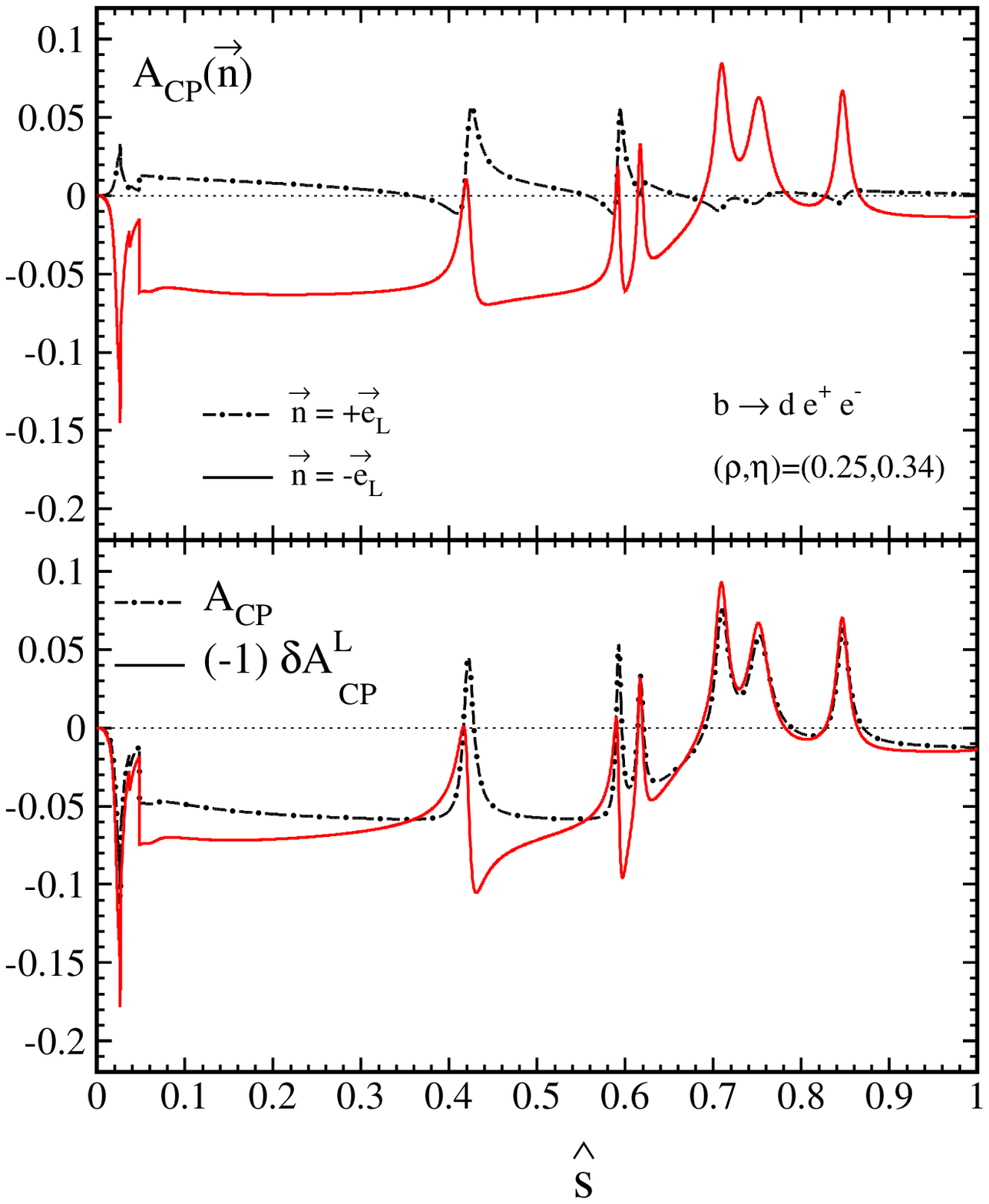,angle=0,width=14cm}
\end{center}
\vspace*{-0.5cm}
\caption{Polarised and unpolarised CP violating rate
asymmetries 
for the decay $b \to d e^+ e^-$
as given in \protect\eqref{eq:acpei}, \protect\eqref{eq:polCPdef}
and \protect\eqref{eq:CPdef}, respectively.}
\label{fig:acp_e} 
\end{figure} 

\begin{figure}[ht]
\begin{center}
\hspace*{-2.0cm}
\epsfig{file=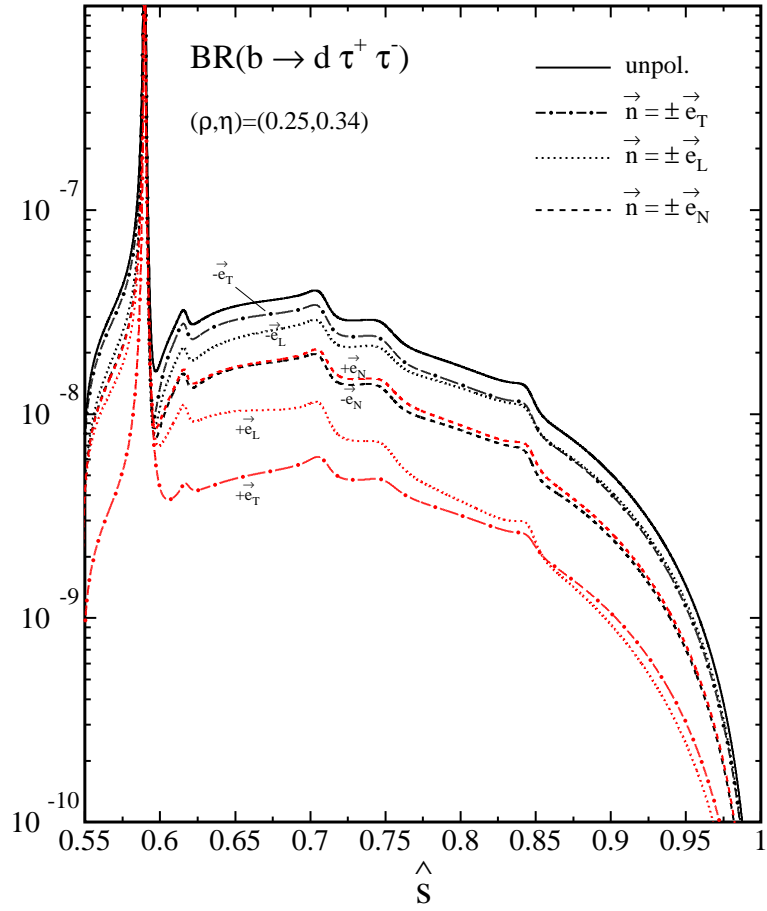,angle=0,width=12cm}
\end{center}
\vspace*{-0.5cm}
\caption{
As in Fig.\ \protect\ref{fig:br_e} for the
decay $b \to d \tau^+ \tau^-$.}
\label{fig:br_tau} 
\end{figure} 

\begin{figure}[ht]
\begin{center}
\hspace*{-2.0cm}
\epsfig{file=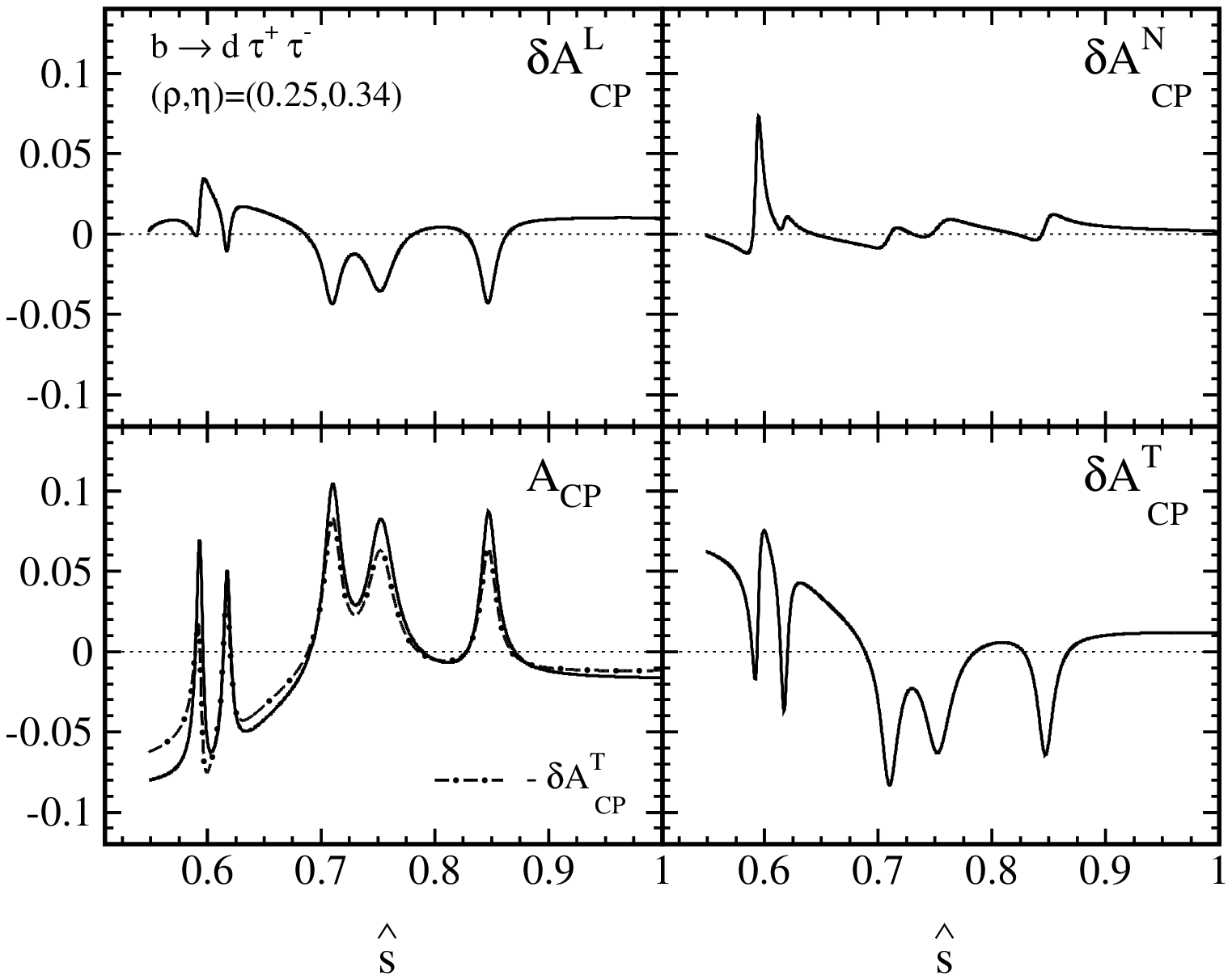,angle=0,width=16cm}
\end{center}
\vspace*{-1.0cm}
\caption{Polarised and unpolarised CP violating rate
asymmetries for the decay $b \to d \tau^+ \tau^-$
as given in \protect\eqref{eq:polCPdef}
and \protect\eqref{eq:CPdef}, respectively.}
\label{fig:acp_tau} 
\end{figure} 

Having given the analytic framework for the various observables, we now turn 
to a discussion of our numerical results. We reiterate that these results are
obtained following the discussions of section \ref{sec:framework}. In 
particular, we take in to account the $\rho-\omega$ resonance structure, 
which to our knowledge has so far been given little importance
except in \cite{Kruger:1997dt}.\\

The basic and essential information
is summarised in figures \ref{fig:br_e} to \ref{fig:acp_tau}. 
In all our numerical estimates, we 
have used the parameters stated in the appendix. 
Furthermore, we have set $m_d = 0$ in our kinematic analysis as its 
dependence in $\Delta(\shat)$ (and similar kinematic factors) is 
negligible. On the other hand, the non-zero value for $m_d$ given
in the appendix
has been used in evaluating the Wilson coupling ${\Cixeff}$. 
We have checked numerically that
the use of small current masses $m_u = 2.3\ \mev$ and 
$m_d = 4.6\ \mev$ in the calculation of ${\Cixeff}$
would result in appreciable changes 
of the asymmetries only in the 
region of the higher $c\bar{c}$ resonances ($\shat > 0.5$) which
is also affected by other theoretical uncertainties; see the discussion below.
The theoretically clean region between the $\rho-\omega$ and
the $J/\Psi$ resonance remains almost unaffected. 
\\

The currently allowed range for the Wolfenstein parameters
\cite{Wolfenstein:1983yz,*Wolfenstein:1964ks}
is given in \cite{Abbaneo:2001bv};
$0.190 < \rho < 0.268$, $0.284 < \eta < 0.366$.
For our analysis we take
$(\rho,\eta)=(0.25,0.34)$.
In terms of the Wolfenstein parameters, $\rho$ and $\eta$,
the parameter $\lu$ is given by the relation,
\be
\lu = \frac{\rho (1-\rho) - \eta^2}{(1-\rho)^2 + \eta^2}
- i \frac{\eta}{(1-\rho)^2 + \eta^2}+ \cdots \quad .
\ee


 In Fig.\ \ref{fig:br_e}, we display the branching ratios for the
decay $b \to d e^+ e^-$ with unpolarized and longitudinally
polarized electrons.
The unpolarized branching ratio (solid line) has been 
obtained with help of  \eqref{eq:BR}.
The corresponding results for polarised final state
leptons have been calculated accordingly using
\eqref{eq:poldecay}.
The dash-dotted and dotted lines corresponds to $\vecn = - \eL$ 
and $\vecn = + \eL$, respectively, where
the unit vector $\eL$ has been defined in \eqref{eq:unitvecdef}.
As can be seen, in the SM, the decay is naturally left-handed and
hence the polarised spectrum for $\vecn = -\eL$ compares with the 
unpolarised spectrum.
Whereas, the polarised $\vecn = \eL$ spectrum is far below the 
unpolarised one; which according to our definition would correspond to a 
{\it wrong sign} decay. In the most relevant kinematical region, between the
$\rho-\omega$ and $c \bar c$ resonances, the branching ratio
is $\sim 3 \times 10^{-7}$. We would like to remind that this is theoretically
the cleanest kinematic bin.\\

In Fig.\ \ref{fig:acp_e}, we present results for the polarised
and unpolarised CP violating rate asymmetries calculated according
to \eqref{eq:acpei}, \eqref{eq:polCPdef}, and \eqref{eq:sigmai}
for the decay $b \to d e^+ e^-$.
We find that the asymmetries for $b \to d \mu^+ \mu^-$
are numerically similar to the results shown here, and hence are not
presented.
As mentioned earlier, only two of the shown four quantities are
linearly independent. 
As can be observed, $\ACP(\vecn = - \eL)$ is much larger than the asymmetry
with opposite lepton polarization implying 
that $\ACP(\vecn = - \eL)$ is quite similar to
the unpolarized asymmetry $\ACP$; the system is naturally polarized.
This can be also seen by the lower half of Fig.\ \ref{fig:acp_e} 
where the unpolarized $\ACP$ and $(-1) \ACPL$ have been plotted.
Here, $\ACP(\vecn = - \eL)$
would be the average of the two curves (lying in the middle between them). 
Note that the polarised CP violating asymmetry $\ACPL$
is comparable and in certain kinematic regions even larger than its 
unpolarized counterpart.
Particularly, in the theoretically clean region, between the 
$\rho$--$\omega$ and the $c\bar c$ resonances, we find $\ACPL$ is about 
$8 \%$ when compared to about $5 \%$ in the 
unpolarized case \cite{Kruger:1997dt,Ali:1998sf}. 
However, in the resonance regions, the polarised asymmetry can reach
values as large as up to $20 \%$ ($\rho$--$\omega$) and 
$11 \%$ ($c\bar c$), respectively.\\

A few other comments are in order.
(i) In our analysis we have utilized the Wolfenstein parameters
$(\rho,\eta) = (0.25,0.34)$ which is in the experimentally 
allowed range given above. However, the results for other values of
these parameters can be easily obtained.
Noticing that since $\Cixeff(\shat)$ only very weakly depends on 
$\rho$ and $\eta$, almost the entire dependence is due the 
prefactors containing the CKM matrix elements; particularly,
in the expressions for the branching ratio and the CP violating 
asymmetries.
In the case of the branching ratio this is the term
$|V_{tb} V_{td}^*/V_{cb}|^2 = \lambda^2 [(1-\rho)^2 + \eta^2]$;
in the case of the CP violating asymmetries it is the factor
$\Imag \lu = -\eta/((1-\rho)^2 + \eta^2)$.
The results for other Wolfenstein parameters can therefore
be obtained by simply rescaling the shown results.
For instance varying $(\rho,\eta)$ in the allowed range
leads to a variation of $|\Imag \lu|$ in the range
$(0.54\div 0.38)$. (For $(\rho,\eta) = (0.25,0.34)$ we find
$|\Imag \lu|=0.5$.)
As can be seen the absolute value of the CP violating asymmetries 
increases with
increasing $\rho$ and $\eta$. On the other hand, the branching
ratio mildly decreases with increasing $\rho$.
(ii) Our results for $\ACP$ in the region of the $\rho - \omega$ resonance are 
smaller than the values found in \cite{Kruger:1997dt}. The discrepancy can 
most probably be traced back on the $\rho-\omega$ inelastic channel
(see the function $G(s)$ in \cite{Kinoshita:1985it}) in which the absorptive 
part should be omitted below the inelastic threshold 
$\sqrt{s} < m_\pi + m_\rho$. On inclusion of the imaginary part also below 
this threshold, we find similarly large contributions at the 
$\rho-\omega$ resonances. 
(iii) In addition, the numerical results in the 
resonance region of the $J/\Psi$ family are subject to theoretical 
uncertainties; usually denoted by a phenomenological factor $\kappa_V$ which
can vary between 2.35 to 1.00. 
For our numerical results we have set $\kappa_V = 1$.
The variation of $\kappa_V$ can affect the width and the
$CP$ asymmetry as has been discussed in \cite{Kruger:1997dt}.\\

The polarised asymmetries, $\ACPN$ and $\ACPT$, are proportional to the 
lepton mass and therefore only relevant in the case of final state tau 
leptons.
In Fig.\ \ref{fig:br_tau}, branching ratios for the
decay $b \to d \tau^+ \tau^-$ are shown for unpolarized (solid line),
longitudinally (dotted), normally (dashed), and transversely 
(dash-dotted) polarized taus. The corresponding branching ratio is 
$\sim \Ord(1 \times 10^{-8})$, requiring a larger luminosity.
With tau leptons as final states, for $\vecn = \pm \eN$; both rates are 
very similar, whereas, for $\vecn = \pm \eT$, the $-\eT$ state is strongly 
favored, as being closer to the unpolarised decay width. Therefore, we would
classify the polarised $\vecn = \eT$ spectra as a {\it wrong sign} decay.\\ 

In Fig.\ \ref{fig:acp_tau}, we show both the polarised
and unpolarised CP violating rate asymmetries calculated according
to \eqref{eq:acpei}, \eqref{eq:polCPdef}, and \eqref{eq:sigmai}
for the decay $b \to d \tau^+ \tau^-$. 
Since $\ACPL$ and $\ACPN$ are small we conclude that
$\ACP(\vecn = +\eL) \simeq \ACP(\vecn = -\eL)$ and
$\ACP(\vecn = +\eN) \simeq \ACP(\vecn = -\eN)$.
On the other hand, $\ACPT$ is comparable to the unpolarised
$\ACP$ as is indicated by the dash-dotted line in the $(2,1)$-panel.
This in turn implies that $\ACP(\vecn = +\eT)$ is very small. As can be
observed, all calculated asymmetries, reach at the maximum about $10 \%$. 

\section{Summary}
\label{sec:summary}
We have made a detailed study of the $CP$ asymmetry for the
process, $b \to d l^+ l^-$, when one of the leptons is in a polarised state.
Our results indicate that when a lepton is in a certain polarised state
$(-\eL,-\eT)$, the decay rates are comparable to the unpolarised 
spectrum. For the case of the normally polarised spectrum, both 
$\pm \eN$ give similar widths but lower than in the case 
of $-\eL$ and $-\eT$. 
In the rest of the polarisation states, which we had defined to be 
the {\it wrong sign} states, the decay rates and the 
corresponding asymmetries
are lower, in comparison to the unpolarised results. In the regions 
which are away from resonance, we find the
polarised $CP$ asymmetries are larger than the corresponding unpolarised 
asymmetry. As expected, the resonance regions show a large
asymmetry and in all of our analysis, we have included the $\rho-\omega$ 
resonance states also. However, as discussed, the results at the resonance 
region suffer from theoretical uncertainties. Barring a small kinematic
window, measuring $\ACP$ requires $d$ quark tagging 
(due to the dominant background $b \to s l^+ l^-$) \cite{Kruger:1997dt}; in 
addition, for the polarised 
asymmetries the polarization needs to be measured. However, 
within the SM given the left-handed nature of the interactions, for the
case of the electron, it is dominantly in the $-\eL$ state. We note that the
results for the electron and muon in a given polarised state do not differ
much and hence measuring a muon in a {\it wrong sign} polarised state $(+\eL)$
can be very sensitive to new physics. Essentially, we need to probe polarised
$(+\eL)$ muons; which we expect to be possible by (i) angular 
distribution of the decay products and (ii) through the life 
time of the $+\eL$ muons, which is enhanced as compared to 
the $-\eL$ state due to 
the dynamics of the SM interaction (left-handed); this is also evident by 
their smaller decay width as observed in Fig.\ \ref{fig:br_e}. 
The situation for the case of the
tau leptons is different and we observe that the $\eT$ polarised state can be 
most sensitive to new physics (see Fig. \ref{fig:br_tau}).
We note that the polarisation observables involve different 
quadratic combinations of the Wilson couplings and hence make for additional 
consistency checks for the unpolarised SM decay spectra and also for a probe 
to new physics which can show in loop effects. In this respect, given a real 
valued $C_{10}$ (for the SM), we 
note that the asymmetries, $\delta\Sigma^{\rm T}$ and $\delta\Sigma^{\rm N}$, 
can be of relevance through the contributions arising from the 
real and imaginary parts of the function $\xi_2$ to the operators, 
$O_1 - O_6$. 

In order to observe a 
3$\sigma$ signal for $A_{CP}(\hat s)$ about $\sim 10^{10}$ $B$ mesons
have been estimated to be required \cite{Kruger:1997dt}.
As already mentioned such a measurement requires a good $d$-quark
tagging to distinguish it from the much more copious decay
$b \to s l^+ l^-$ and hence
will be a challenging task at
future hadronic colliders like LHCb, BTeV, ATLAS or CMS 
\cite{Harnew:1999sq}. 
More dedicated experiments like Super-BABAR and Super-BELLE
should be able to achieve this goal.
BELLE and BABAR have already measured the rare decay $b \to s l^+ l^-$
which could be measured with great accuracy at these high luminosity
upgrades.
Given enough statistics,
and excellent kaon/pion identification, they may be able to 
measure $b \to d l^+ l^-$ or the exclusive process 
$B \to \rho l^+ l^-$.\footnote{We thank A.\ Ali for pointing this
out to us.}
In contrast, 
for measuring $\ACP(\vecn = - \eL)$ we need a similar number
of produced $B$-mesons, provided an efficient polarisation measurement
is possible,
since the branching rates are very much alike as discussed previously. 
In this regard, given the almost massless nature of the electrons, the
$CP$ asymmetry for the decay $b \to d e^+ e^-$ would correspond to a 
polarised asymmetry as in (\ref{eq:acpei}). As a result, if our SM 
expectations for the asymmetry as calculated here (see
Fig. \ref{fig:acp_e}) are not met with, we might have a reasonable chance
to look for new physics in polarised decays. In this paper, we 
have classified
all such polarised decays as {\it wrong sign} decays and
these could be viable modes in search for new physics.
\begin{center}
{\bf Acknowledgments} 
\end{center} 
The work of Balaji has been supported by the
Bundesministerium f\"ur Bildung, Wissenschaft, Forschung und Technologie,
Bonn under contract no. 05HT1PEA9. 
Part of the work of I.\ S.\ has been done at the University of Dortmund
supported by the 'Bundesministerium f\"ur Bildung und Forschung', 
Berlin/Bonn. 
We thank Frank Kr\"uger and
Lalit Sehgal for many useful discussions. 
I.\ S.\ thanks Matthias Steinhauser for helpful comments on the
input parameters. 
We also thank E.\ Reya and A.\ Ali for comments and suggestions.


\begin{appendix} 
\section{Input Parameters}
\begin{equation*}
\begin{gathered}
m_b = 4.8\ \gev, m_t = 165\ \gev, m_c = 1.3\ \gev,
\\
m_u = m_d = m_\pi = 140\ \mev, 
\\
m_e = 0.511\ \mev, m_\mu = 0.106\ \gev, m_\tau = 1.777\ \gev
\\
\mu = m_b, B(B \to X_c e \bar \nu_e)= 10.4 \%
\\
M_W = 80.423\ \gev, M_Z = 91.1876\ \gev, \sin^2(\theta_W) = 0.23,
\\
\alpha_{\rm em} = 1/129,
\alpha_s(M_Z) = 0.1183, \Lambda^{(5)}_\msbar = 230\ \mev
\\
\lambda = 0.2237, A = 0.8113, \rho = 0.25, \eta = 0.34
\end{gathered}
\end{equation*}
\end{appendix}

\clearpage
\bibliographystyle{/afs/desy.de/user/s/schien/Bibliography/test}
\bibliography{/afs/desy.de/user/s/schien/Bibliography/bphysics}

\end{document}